\documentclass[aps,prl,twocolumn, superscriptaddress, amsmath]{revtex4-1}

\usepackage[usenames,dvipsnames]{color}
\usepackage{graphicx}
\usepackage{dcolumn}
\usepackage{simplewick}
\usepackage{bm}
\usepackage{amsmath}
\usepackage{braket}
\usepackage{float}
\usepackage{threeparttable, tablefootnote}
\usepackage{hyperref}
\usepackage{natbib}
\usepackage{multirow}
\usepackage{adjustbox}
\usepackage{hyperref}
\hypersetup{colorlinks=true, urlcolor=blue, citecolor=blue}

\newcommand{\pzo}{{\rm PbZrO$_3$}}
\newcommand{\cgb}{{\rm CsGeBr$_3$}}
\newcommand{\pto}{{\rm PbTiO$_3$}}
\newcommand{\pho}{{\rm PbHfO$_3$}}
\newcommand{\hfo}{{\rm HfO$_2$}}
\newcommand{\aln}{{\rm AlN}}
\begin{document}

\title{Relaxation approach to  quantum-mechanical  modeling of ferroelectric and antiferroelectric phase transitions}

\author{Nikhilesh Maity}
\email{nikhileshm@usf.edu}
\affiliation{Department of Physics, University of South Florida, Tampa, Florida 33620, USA}
\author{Sergey Lisenkov}
\affiliation{Department of Physics, University of South Florida, Tampa, Florida 33620, USA}
\author{Arlies Valdespino}
\affiliation{Department of Physics, University of South Florida, Tampa, Florida 33620, USA}
\author{Milan Haddad}
\affiliation{School of Materials Science and Engineering, Georgia Institute of Technology, Atlanta, GA 30318, USA}
\author{Lewys Jones}
\affiliation{School of Physics, Trinity College Dublin, Dublin, Ireland}
\affiliation{Advanced Microscopy Laboratory, Centre for Research on Adaptive Nanostructures and Nanodevices (CRANN), Dublin 2, Ireland}
\author{Amit Kumar}
\affiliation{Centre for Quantum Materials and Technologies, School of Mathematics and Physics, Queen's University Belfast, Belfast, UK}
\author{Nazanin Bassiri-Gharb}
\affiliation{Advanced Research Institute, Georgia Institute of Technology, Atlanta, GA 30313, USA}
\affiliation{G.W. Woodruff School of Mechanical Engineering, School of Materials Science and Engineering, Georgia Institute of Technology, Atlanta, GA 30332, USA}
\author{Inna Ponomareva}
\email{iponomar@usf.edu}
\affiliation{Department of Physics, University of South Florida, Tampa, Florida 33620, USA}

\date{\today}

\begin{abstract}
Ferroelectrics and antiferroelectrics are the electric counterparts of ferromagnets and antiferromagnets. These materials undergo temperature- and electric-field–induced phase transitions that give rise to their characteristic hysteresis loops. Modeling such hysteresis loops and associated phase transitions enables both a deeper fundamental understanding and reliable property predictions for this important class of materials. To date, modeling has largely relied on classical approaches, often remaining qualitative and/or empirical. Traditional interpretation of these transitions rests on two assumptions: (i) they are activated Arrhenius-type processes and (ii) they occur well within the classical regime. Here, we demonstrate that a model can instead be built on two ``orthogonal" assumptions: (i) the phase transitions are relaxational processes and (ii) they require a quantum mechanical treatment. Applying this model to both antiferroelectrics and ferroelectrics overcomes the limitations of traditional models and enables efficient first-principles simulations of phase transitions.  The success of our unconventional approach highlights the significance of quantum mechanics in transitions long regarded as purely classical. We anticipate that this framework will be applicable to a broad range of phase transitions, including magnetic, elastic, multiferroic, and electronic, along with modeling of quantum tunneling, rates of chemical reactions, and others.

\end{abstract}

\maketitle
\begin{figure*}
\centering
\includegraphics[width=1.0\textwidth]{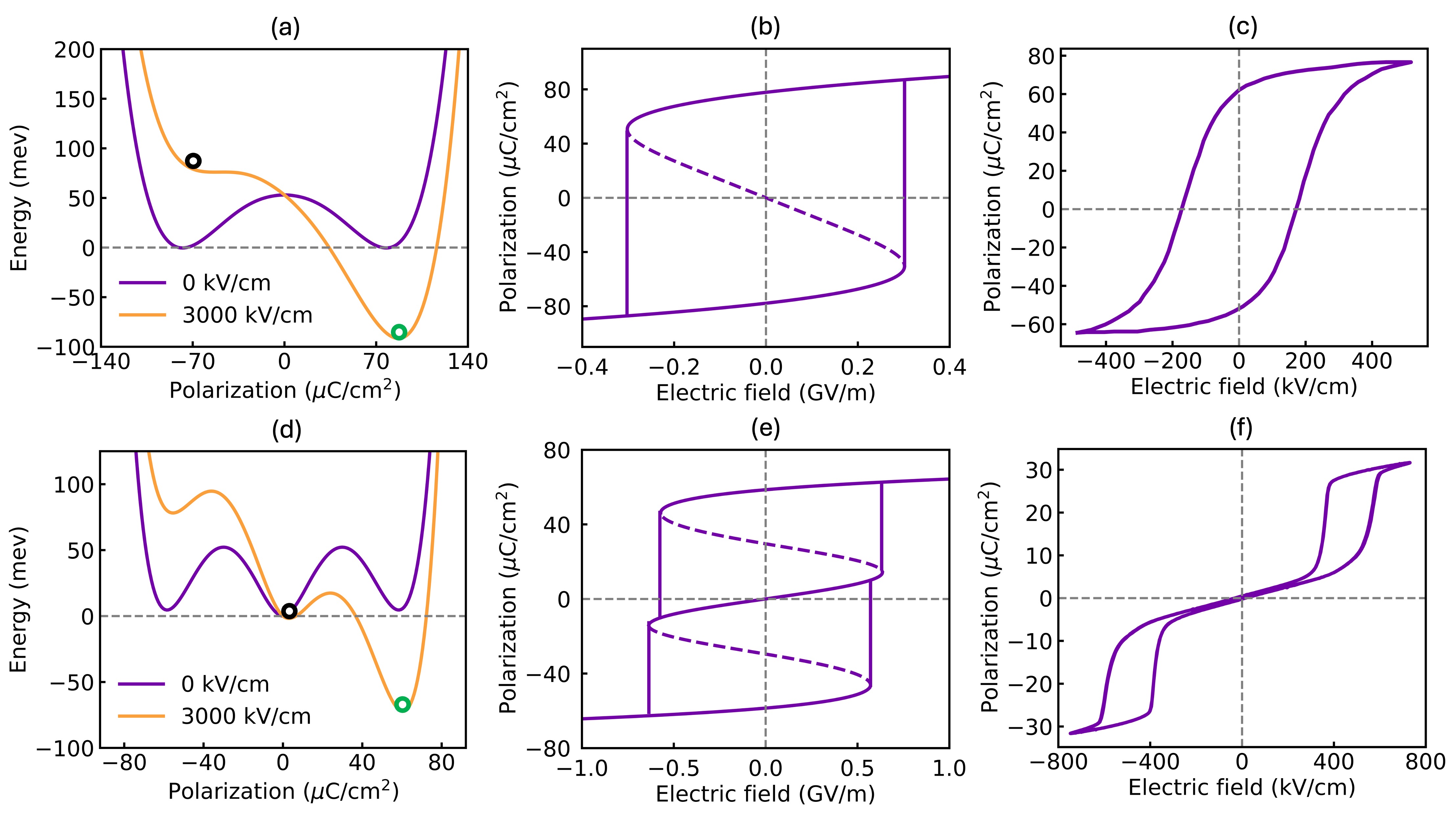}
\caption{Energy per formula unit as a function of polarization for ferroelectric \pto\ (a) and antiferroelectric \pzo\ (d) as computed from DFT simulations.  Equation of state for  \pto\ (b) and \pzo\ (e) obtained from their supercell energies. Black and green dots indicate metastable state and ground state, respectively. The experimental RT polarization as a function of electric field for \pto\  from Ref.~\onlinecite{shin2022imprinted} and for \pzo\ measured in this work (f). For \pzo, the electric field is applied along [111] pseudo-cubic direction.}
\label{fig1}
\end{figure*}

Ferroelectrics and antiferroelectrics are the electric counterparts for ferromagnets and antiferromagnets. They are important class of materials both from fundamental science and applied perspectives. For example, ferroelectrics are used in Memory (FeRAM), sensors, actuators, tunable capacitors, energy harvesters~\cite{mikolajick2021next,troiler2020impact,mikolajick2020past}, while antiferroelectrics are utilized in high-energy-density capacitors, pulse power, electrocaloric cooling, and advanced thermal
switches~\cite{randall2021antiferroelectrics,liu2018antiferroelectrics,pirc2014negative,liu2023low}. Ferroelectric and antiferroelectric phase transitions originate from the multiple well free energy profile. As shown in Fig. \ref{fig1}(a) and (d), the zero-Kelvin free-energy, $U$,  as a function of polarization, $P$, exhibits a double-well form for the ferroelectric phase and a triple-well form for the antiferroelectric phase. The energies are computed from Density Functional Theory (DFT) simulations. 
Applied electric field ``tilts" the energy profile (yellow line in Fig.\ref{fig1}) so that one of the minima becomes energetically more favorable (stable state)  and the system will eventually transition into it. If the systems initially was located in the other minima (metastable state, indicated by the black point in Fig.\ref{fig1}), it has to overcome the energy barrier.  It is typically assumed that this is an Arrhenius type process, which describes transitions over barriers significantly higher than $k_B T$, where $k_B$ is the Boltzmann constant and $T$ is the temperature. This is an activated process since transition rate is proportional to $e^{-\mathcal{E}_A/k_B T}$, where $\mathcal{E}_A$ is the activation energy, which is typically very close to the height of the  energy barrier. As a result, if the barrier is very high the transitions rates are very low, giving rise to extremely long lifetime of the metastable states. 

Figure~\ref{fig1} (d) shows an energy landscape computed from DFT simulations for prototypical antiferroelectric \pzo. The transition rate  can be estimated by  $\nu e^{-\mathcal{E}_A/k_B T}$, where $\nu$ is the attempt frequency and can be approximated with the soft mode frequency~\cite{kashikar2025dftbased}. For the typical soft mode frequency of 1~THz, the transition rate at room temperature and the experimental forward-switching field of 315~kV/cm  is 2.0 $\times$ 10$^{-40}$ s$^{-1}$, which is unphysical. As the field increases, the barrier $\Delta \mathcal{E}_A$ decreases and the rates go up. 

One way to quantify the fields required to overcome the barrier is through converting the energy $U(P)$ in Fig.~\ref{fig1} (a) and (d) into the equation of state as follows~\cite{alma99379576459406599}: the free  energy in the presence of the electric field, $E$, is $F(P)=U(P)-EPV$, where $V$ is the  volume. Setting the derivative $\frac{\partial F}{\partial P}$ to zero produces the equation of state, $P(E)$, shown by  dashed line in Fig.~\ref{fig1}(b) and (e). Avoiding the thermodynamically unstable branches  associated with negative susceptibility $\chi = \frac{\partial^2 F}{\partial P^2}$ results in the loops shown by solid lines in Fig.~\ref{fig1}(b) and (e). 
The model predicts that at zero Kelvin, \pzo\ exhibits ``mixed'' behavior, characterized by antiferroelectric-like double hysteresis loops together with a nonzero spontaneous polarization. In contrast, the experimentally measured hysteresis loop, shown in Fig.~\ref{fig1}(f), does not exhibit this mixed behavior. The measurement was performed on a highly 042-oriented \pzo\ thin film (approximately 290$\pm$5 nm thick), chemical solution processed on platinized silicon, under previously reported conditions~\cite{haddad2025chemical}.
For ferroelectrics (see Fig.~\ref{fig1}(b)) we find coercive fields of $\sim$ 3018 kV/cm, which  exceeds by order of magnitude the experimental ones ~\cite{shin2022imprinted} ($\sim$ 175 kV/cm) (see Fig.~\ref{fig1}(c)). We elaborate that within the equation of state model the system transitions from the metastable phase to the stable one once the local minimum aligns with the barrier top, which corresponds to infinitely long lifetime of the metastable phase. Although the model does not give access to time or frequency dependence, infinitely long lifetime of metastable phase effectively models extremely high frequency of AC electric field so that the system never has time to experience the fluctuation needed to overcome the barrier as per Arrhenius process.  This explains unphysically high coercive/switching fields.  In Ref.~\cite{ghosh2022unusual} this model was augmented with the frequency- and temperature-dependent transition rates, but the application of the model here still predicted coercive field that are  too high, suggesting that traditional transition state-based models~\cite{shol-book} may not be fully adequate to describe such phase transitions.

Traditionally, the model failures are attributed to the inability of present DFT calculations to adequately capture the transition states and associated barriers~\cite{esswein2022ferroelectric,paul2017accuracy}. Indeed, in DFT simulations we are limited to small simulations supercells, which critically limits the phase space available to the system. For example, configurations with domains are unlikely to appear, although they have been shown to play a critical role in phase transitions~\cite{liu2016intrinsic}. The other possible culprit is the pristine nature of the material in simulations, while real materials have defects and surfaces that are expected to reduce the barriers for transitions. 

So far, success in modeling of ferroelectric and antiferroelectric phase transitions has been achieved with Landau-type  approaches~\cite{toledano2016theory,hatt2000landau}, related phase-field modeling~\cite{masuda2025atomic,fan2023phase}, first-principles-based effective Hamiltonians~\cite{bin2016wang,kashikar2024coexistence,mani2015finite}, shell models~\cite{goncalves2017finite,graf2014phase}, and machine-learned potentials~\cite{gigli2024modeling,zhang2024finite}.  These include predictions of phase transition temperatures, hysteresis loops, dynamics, temperature evolutions of polarizations among others. The common theme between them is that they all utilize classical frameworks. Nevertheless, the counterintuitive possibility that the classical treatment may not be sufficient for ferrolectrics has been raised in a few pioneering studies. Using path integral technique, Zhong and Vandebilt showed that zero point energy is sufficient to destroy ferroelectric ordering in quantum paraelectric SrTiO$_3$ \cite{PhysRevB.53.5047}.  Subsequently, Geneste et al. applied path-integral molecular dynamics to BaTiO$_3$ to reveal that in contrast to classical picture this ferroelectric exhibits strong anharmonicity down to lowest temperatures, which results in  enhanced the dielectric and piezoelectric responses \cite{PhysRevB.87.014113}. Through the same path integral approach it was found that quantum effects lead to a significant reduction of transition temperatures (up to 50~K) in BaTiO$_3$, which revealed that quantum effects play significant role even at temperatures as high as room temperature \cite{dammak2018nuclear}.  In methodologically different work, Esswein and Spaldin demonstrated that the quantum effects are responsible for classifying ferroic as ferroelectric, paraelectric, and quantum paraelectric  can be captured through single-particle Schrodinger equation with the DFT-calculated potential  \cite{esswein2022ferroelectric}. However, it has not been established whether these techniques are able to resolve hysteresis loops controversies outlined above. 

Here, we introduce a fundamentally different first-principles framework for (anti)ferroelectricity. Instead of treating the phase transition classically and as an Arrhenius-type activated process, we describe it quantum mechanically and as a relaxation-driven evolution toward equilibrium. This unconventional framework (semi-)quantitatively reproduces ferroelectric and antiferroelectric hysteresis loops across a broad range of materials, providing strong evidence that quantum mechanics and relaxation dynamics are essential ingredients of these phase transitions.

{\it Ground-State Relaxation (GSR) Model.} Let us begin with describing the system being in a quantum state $\ket{\psi}$ that propagates in time, that is $\ket{\psi(t)}$. For time-independent Hamiltonian $H=\frac{\mathcal{P}^2}{2M}+U(P)$, where $\mathcal{P}$ is the momentum, this can be achieved as $\ket{\psi(t)}=\sum_n c_n(0)e^{-i\frac{ E_n}{\hbar}t}\ket{\phi_n}$, where $c_n(0)=\braket{\phi_n|\psi(t=0)}$, while $\ket{\phi_n}$ and $E_n$ are the eigenstates and eigenvalues  of the Hamiltonian, respectively. For a given Hamiltonian $\ket{\phi_n}$ and $E_n$ can be evaluated numerically. Suppose $\ket{\psi(t=0)}$ corresponds to the metastable state of the system (black dot in Fig.~\ref{fig1}(a) and (d)). Then at time $t$, the probability of finding the system in the state $\ket{\phi_n}$ is $\lvert\braket{\phi_n|\psi(t)}\rvert^2=|c_n(0)|^2$ and  independent of time.  This means that the system will never settle into the ground state, which is the stable state associated with the green dot in Fig.~\ref{fig1}(a) and (d). However, any real system will relax from excited to the ground state (black and green dots in  in Fig.\ref{fig1}(a) and (d), respectively), owing to the interaction with the environment. This can be incorporated into the model by replacing $E_n$ in $\ket{\psi(t)}$ with $E_n'=E_n-i\hbar \frac{\gamma_{n}}{2}$, where $\gamma_{n} = 1/\tau_{n}$ is the relaxation rate, while $\tau_{n}$ is the  associated relaxation time from the excited state $\ket{\phi_n}$ into the other states ~\cite{cohen2019quantum}. We will later discuss how relaxation rates can be computed but for now we take them to be  $\gamma_n=k_{scale}\sum_{E_m<E_n}
\frac{E_n - E_m}{h}$.  The proportionality constant $k_{scale}$ is taken here to be 4.56$\times$10$^{-2}$ and will be justified later. This expression implies that the relaxation rate from the state $\ket{\phi_n}$ is proportional to the sum of Bohr frequencies associated with all the states below $\ket{\phi_n}$ in energy.
Now the probability of finding the system in the excited state $\ket{\phi_n}$ is $|c_n(0)|^2e^{-\gamma_{n}t}$ and decays exponentially with time. Note, that  such a decay is not norm-conserving and, therefore, we find it convenient to renormalize the state vector $\ket{\psi(t)}$. The polarization can be computed as the expectation value of the polarization operator $P(t)=\braket{\psi(t)|\hat{P}|\psi(t)}$.

\begin{figure*}
\centering
\includegraphics[width=1.0\textwidth]{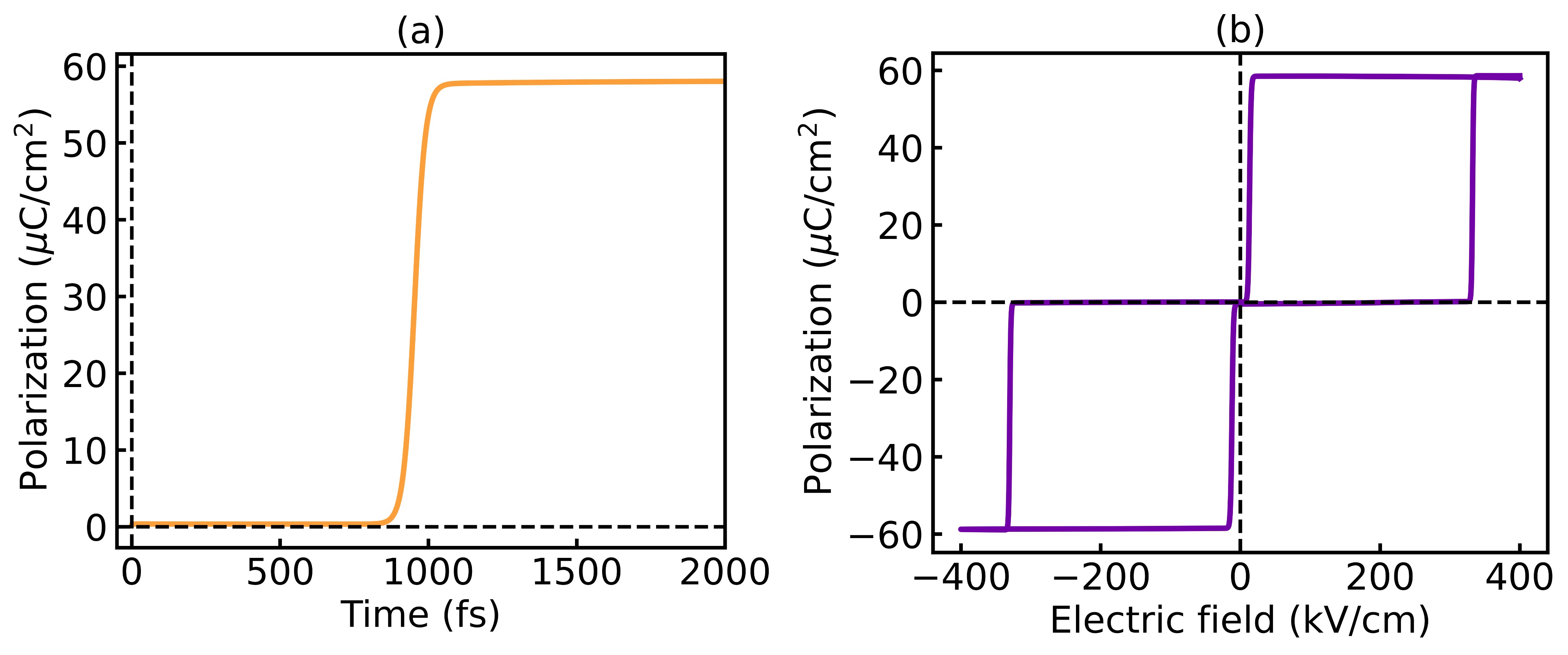}
\caption{(a) Time evolution of the polarization in antiferroelectric \pzo\ during the transition from the metastable state to the ground state at 0 K under a DC electric field of 400 kV/cm. (b) Hysteresis loops of \pzo\ computed using the GSR approach. The simulations were performed using an AC field frequency of $\nu=0.017$ THz and an integration time step of $\Delta t=1$ fs.}
\label{fig2}
\end{figure*}
Figure \ref{fig2}(a) shows how the system that started in the metastable state of \pzo\,  transitions  into the ground state  at the field of 400~kV/cm applied along the polar direction. We note that  the time it takes to transition is defined by the relaxation time we used in the modeling. Since the one we used is for illustrative purposes only, the transition time is not necessarily physical. This is, however,  improvable as the relaxation times could be obtained, at least in principle, from DFT or based on the experimental input.  The metastable state was modeled as a gaussian wave packet centered at $P=$0 $\mu$C/cm$^2$ and 0.08 $\mu$C/cm$^2$ wide. Note, that we choose to focus on relaxation part of the process only and, therefore, turn the intrinsic dynamics off by setting Bohr frequencies to zero.

Next we want to extend these ideas to a time-dependent Hamiltonian $H=\frac{\mathcal{P}^2}{2M}+U(P)-PE(t)V$, where $E(t)$ is now the time dependent electric field. We can discretize the simulation time into intervals and keep the field constant during each interval, while updating it between intervals. After the field is updated we project the state $\ket{\psi(t)}$ onto the new basis associated with the eigensatates of the $H$ for the given field and renormalize it again to account for the numerical drift due to potential incompleteness of basis. Note, that we use 80 lowest energy eigenstates for the basis, which we find sufficient to represent the state of the system. We simulate 1.25 periods of AC field applied along polar direction of the rhombohedral phase of \pzo\ and show the results in Fig.~\ref{fig2}(b). The first quarter is removed for presentation purposes. The model correctly predicts antiferroelectric hysteresis loops with switching fields comparable to experimental values (see Fig.~\ref{fig1} (f)), despite the calculations corresponding to 0 K. We find that the relaxation approach is capable of both reproducing an antiferroelectric hysteresis loop and predicting switching fields in good agreement with experiment, in contrast to the models discussed in the introduction. However, the GSR approach is limited to zero Kelvin. To overcome this limitation we turn to the density matrix approach with the major advantage that it allows for natural incorporation of temperature. 

{\it Density Matrix Based (DMB) Model. } The equilibrium state of a system is now described by the density operator $\rho^{eq}=\frac{1}{Z}\sum_n e^{-\frac{E_n} {k_B T}}\ket{\phi_n}\bra{\phi_n}$, where $Z$ is the canonical partition function. In such an approach the state of the system is modeled by the density operator $\rho(t)=\ket{\psi(t)}\bra{\psi(t)}$ whose time evolution is given by the  Liouville equation \cite{alma99379890528406599} $\frac{d \rho}{dt} = -\frac{i}{\hbar}[H,\rho]$. The equation, however, does not include the relaxation term. 
To recover the predictions of the GSR model we use the same expansion for $\ket{\psi(t)}$ in the $\ket{\phi_n}$ basis as before,  and compute $d\rho/dt$. This approach contributes a relaxation term into the Liouville equation (written in $\ket{\phi_n}$ basis) as follows
\begin{equation}
    \frac{d \rho_{nm}}{dt} =   -\frac{i}{\hbar}[H,\rho]_{nm} - \gamma_{nm}\rho_{nm},
    \label{rho_t}
\end{equation}
where $\gamma_{nm}=(\gamma_n+\gamma_m)/2$. 
The first term on the right hand side describes the coherent dynamics, while the last one models decay of densities (both populations and coherences) due to transitions into the ground state. Just as before, the equation does not capture the gain in densities and, therefore, does not conserve the trace of the density matrix, which could be remedied by renormalization of the matrix by its trace.  The polarization can be computed as the expectation value for the polarization  operator  $P=Tr\{\rho \hat{P}\}$. This approach reproduces  $P(t)$ and $P(E)$ dependencies computed with the GSR approach shown in Fig.~\ref{fig2}. 

To incorporate temperature  we recall that at finite temperature the system will be relaxing to $\rho^{eq}$ rather than the ground state density implied by Eq.(\ref{rho_t}). This requirement can be captured by updating  Eq.(\ref{rho_t}) as follows 
\begin{equation}
\frac{d \rho_{nm}}{dt} =   -\frac{i}{\hbar}[H,\rho]_{nm} - \gamma_{nm}(\rho_{nm}-\delta_{nm}\rho_{nm}^{eq})
\label{rho_t_eq}
\end{equation}
where $\delta_{nm}$ is the Kronecker delta. The populations, $\rho_{nn}$, can now both decrease and increase  owing to the second term in parenthesis  to achieve their equilibrium values. Following approach of Ref.\cite{alma99379890528406599} we ensure that the transition rate from the lower energy $E_n$ to the higher energy state $E_m$ is normalized by the probability factor $e^{-(E_m-E_n)/k_BT}$. 
 The relaxation rate for the state $\ket{\psi(t)}$  is now $\gamma_n=k_{scale}
\left [\sum_{E_m<E_n}
\frac{E_n - E_m}{h} +\sum_{E_m>E_n}
\frac{E_m - E_n}{h} e^{-\frac{E_m-E_n}{k_BT} }\right ]$. There exists  analytical solution for the relaxation term: $\rho_{nm}^{rlx}(t) = \rho_{nm}^{eq}(1-e^{- \gamma_{nm}t}) + \rho_{nm}(0)  e^{- \gamma_{nm}t}$. Finally, we augment Eq.(\ref{rho_t_eq}) with a term  that allows to conserve trace of the density matrix (see Supplementary Materials (SM) for derivation)
\begin{equation}
\frac{d \rho_{nm}}{dt} =   -\frac{i}{\hbar}[H,\rho]_{nm} - \gamma_{nm}(\rho_{nm}-\delta_{nm}\rho_{nm}^{eq})+ \rho_{nm} \sum_j\gamma_{j} (\rho_{jj}-\rho_{jj}^{eq})
\label{rho_t_eq_trace}
\end{equation}

\begin{figure*}
\centering
\includegraphics[width=1.0\textwidth]{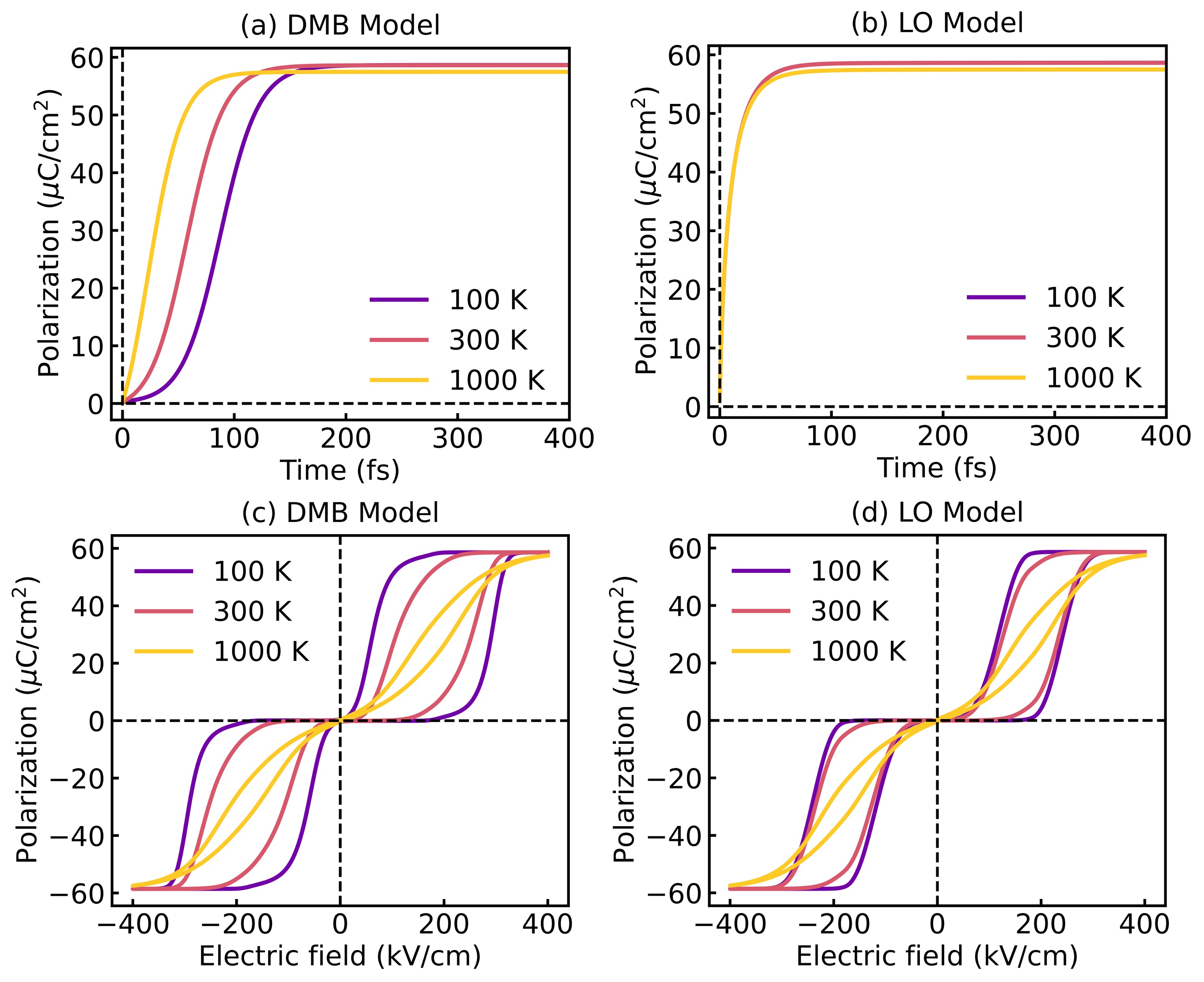}
\caption{(a)-(b) Time evolution of the polarization in \pzo\ during the transition from the metastable state to the stable state under a DC electric field of 400 kV/cm at different temperatures, computed using the models specified in the panel titles. (c)-(d) Polarization as a function of electric field in \pzo\ at different temperatures, computed using the models specified in the panel titles. The simulations were performed using an AC field frequency of $\nu=0.17$ THz and an integration time step of $\Delta t=0.1$ fs.}
\label{fig3}
\end{figure*}

Figure ~\ref{fig3}(a) predicts polarization evolution during relaxation from the metastable state of \pzo\, into a stable one computed within the DMB model for  different temperatures. At 0.1 K it agrees well with the one from relaxation to the GSR model. At higher temperatures relaxation occurs much faster  owing to the  temperature dependent excitations rates in Eq.(\ref{rho_t_eq}). The $P(E)$ loops were computed in the temperature range of 10~K to 1000~K and are given in Fig.~S2(a). Representative data are shown in Fig.~\ref{fig3}(c) and demonstrate that the loops slim down as temperature increases, capturing the temperature evolution from antiferroelectric to paraelectric behavior. Note, that for all the data presented so far we turn off intrinsic dynamics, as it is expected to average to zero. Later on we comment on its effects.

{\it Lindblad Operators (LO) Models.} To provide further grounds to the relaxation approach for quantum-mechanical modeling of antiferroelectrics we implement it within Lindblad formalism. We begin with  stating the master equations, that govern dynamics in the presence of relaxational processes, in terms of Lindblad operators $L_{jk}$~\cite{nielsen2010quantum}, which represent interaction of the system with the environment:
\begin{equation}
\frac{d \rho}{dt} = -\frac{i}{\hbar}[H,\rho]+\sum_{kj}[  2L_{kj}\rho L_{kj}^\dagger - \{ L_{kj}^\dagger L_{kj},\rho\}  ]
\label{lindblad}
\end{equation}
where curly brackets indicate anticommutator. The Lindblad operator, $L_{kj}=\sqrt{\gamma_{kj}}\ket{\phi_k}\bra{\phi_j}$, describes transition from state $\ket{\phi_j}$ into state $\ket{\phi_k}$ caused by the interaction with the environment, at a rate  given by $\gamma_{kj}$. The summations go over all eigenstates. We use the same transition rates as for DMB approach, namely for the down-transitions ($E_k<E_j$) the rate is $\gamma_{kj}=k_{scale}|E_j-E_k|/h $, while for the up-transitions the rate is $\gamma_{jk}=k_{scale}e^{-\frac{E_j-E_k}{k_BT}}|E_j-E_k|/h $. The Lindblad approach preserves trace of the density matrix so we do not have to renormalize it. 
 In the basis of $\ket{\phi_n}$ the master equation (\ref{lindblad}) becomes
 \begin{equation}
\frac{d \rho_{nm}}{dt} = -\frac{i}{\hbar}[H,\rho]_{nm}+
2\delta_{nm} \sum_j \gamma_{nj}\rho_{jj} 
- \rho_{nm}\sum_k(\gamma_{kn}+\gamma_{km})
\label{lindblad_matrix}
\end{equation}
The second term on the right hand side  describes the density in-flow into populations $\rho_{nn}$  due to transitions from all the other states, while the third term describes the density out-flow (from both populations and coherences, $\rho_{nm}$) due to transitions to all the other states. Comparing this equation with Eq.(\ref{rho_t_eq_trace}) we note that the overall rate of population decay from a given state is twice the one in DMB model. We  bring the two models in closer correspondence by rescaling the rates in DMB model by a factor of 2. We note that LO model preserves detailed balance. Indeed, let us consider the change in $\rho_{nn}$ due to relaxation and excitations to state $\ket{\phi_m}$. From Eq.(\ref{lindblad_matrix}) we obtain $\frac{d \rho_{nn}}{dt} = 2\gamma_{nm}\rho_{mm}- 2 \gamma_{mn}\rho_{nn}$. 
At equilibrium $\frac{d \rho_{nn}}{dt} = $ 0 so we have   
$\gamma_{nm}\rho_{mm} = \gamma_{mn}\rho_{nn}$, which is the condition of detailed balance. Furthermore, $\gamma_{nm}/\gamma_{mn} = \rho_{nn}/\rho_{mm}=e^{-\frac{E_n-E_m}{k_BT}}$, which justifies our choice for the relaxation rates in LO model. 
Let us contrast this to previous models. From Eq.(\ref{rho_t_eq_trace}) we obtain $\frac{d\rho_{nn}}{dt}=-\gamma_{nn}(\rho_{nn}-\rho_{nn}^{eq})+\rho_{nn} \sum_j \gamma_{j}(\rho_{jj}-\rho_{jj}^{eq}) $. The sum in the last term is the same for all populations. Therefore, the rate of population change  due to normalization is proportional to the population. The equation does not have an explicit dependence on density matrix elements for any other state, unlike LO approach. So it is clear that the detailed balance is not satisfied. This equation is better interpreted as flow of populations to and from heat reservoir. If we now apply this to relaxation to the ground state models we get  $\frac{d\rho_{nn}}{dt}=-\gamma_{nn}\rho_{nn}+\rho_{nn} \sum_j \gamma_{j}\rho_{jj} $. Since both $\gamma_{j}$ and $\rho_{nn}$ are nonnegative we can easily see that the first term on the right hand side describes the rate of population loss, while the second one describes the rate of population gain.
 
Figure \ref{fig3}(b) and (d) give prediction from the LO Model. We find that the shape of the hysteresis loops change significantly with respect to the previous approaches. In particular, the low temperature loops are slanted rather than square in a better agreement with experimental data in case of \pzo\, (see Fig.~\ref{fig1}(d)). 

On the basis of our data, we conclude that all models developed here are capable of describing  antiferroelectric behavior with switching fields comparable to experimental ones. This result provides strong evidence that (anti)ferroelectric phase transitions can be well modeled as relaxational processes within quantum mechanical framework. Furthermore, DMB and LO models give access to finite-temperature predictions.   
The  methodological difference between the  models is in the way they model dissipative terms. The GSR model accomplishes that through decay of individual states followed by subsequent renormalization of the state vector. In DMB approach the decay and repopulation of the states is achieved through implicit exchange with the thermal bath. In the LO model, a decrease in the population of one state is accompanied by an equal increase in the population of another state, thereby preserving detailed balance. We believe that these difference result in the differences in predictions of $P(t)$ and $P(E)$ evolutions given in Fig.~\ref{fig3}. The DMB approach predicts step-like $P(t)$ evolution at low temperatures consistent with some experimental data~\cite{pantel2010switching,schutrumpf2012polarization}, while LO-based models predict exponential evolution. Consequently, DMB model predicts square hysteresis loops in contrast to slanted ones from LO model. 

{\it Discussion of the models parameters,  approximations, and limitations}. Our models require the energy as a function of polarization along the distortion path (see Fig. \ref{fig1} and S7), which is computed here using DFT. The distortion path itself is generated by linear interpolation between two or more phases, as described in the Technical Details of the DFT Simulations section. Such linear interpolation corresponds to a homogeneous polarization reversal.
We argue that this homogeneous polarization reversal represents the nucleation of a nanodomain of the stable phase within the matrix of the metastable phase. Within the well-established nucleation-limited-switching model \cite{PhysRevB.66.214109}, the switching dynamics are dominated by the time required for domain nucleation, whereas subsequent domain propagation occurs on a much shorter timescale. We therefore believe that our model accurately captures the rate-limiting process governing phase switching, which explains its predictive success.
In the case of \pzo, the supercell volume is 4.4~nm$^3$, which is sufficiently large to represent a realistic nanoscale nucleus.

The model also requires the mass associated with the polar mode, which enters the Hamiltonian as a  parameter. This mass can be determined from DFT calculations. In particular, the harmonic frequency at the bottom of the potential well satisfies $
\omega^2=\frac{1}{M}\left\langle \phi_0 \left| \frac{d^2U}{dx^2} \right| \phi_0 \right\rangle,
$
where $|\phi_0\rangle$ is the ground-state wavefunction. In this work, we determine the mass using the frequency of the polar mode computed from DFT. Equivalently, the mass can be chosen such that, within the harmonic approximation, the energy spacing between the ground and first excited states satisfies $\Delta E^{\mathrm{harm}}=\hbar\omega$.
The mass scales with the size of the simulation supercell and controls both the energy-level spacing and the spatial localization of the eigenstates. Larger masses lead to more closely spaced energy levels and more localized eigenstates, whereas smaller masses produce the opposite behavior. Consequently, increasing the supercell size drives the system from a quantum-mechanical regime toward a  classical one.
The model additionally requires the Born effective charge, $Z^*$, which is used to convert polarization into the corresponding supercell distortion according to $ x=\frac{PV}{Z^*}$.
The Born effective charge can be readily obtained from DFT calculations (see SM).

The remaining model parameter is the relaxation rates, $\gamma_{nm}$, which are taken here to be proportional to Bohr frequencies.  Relaxation rates for an open quantum system connected to a heat bath  could be computed using the ``golden rule" for the transition rates  \cite{alma99379890528406599}
\begin{equation}
\gamma_{kj} = \frac{2\pi}{\hbar Z} \sum_{NN'} |\langle kN | \mathcal{V} | jN' \rangle|^2 e^{- E_{N'}/k_BT} \delta(E_{N'} - E_N - \hbar\omega_{kj})
\end{equation}
where $N$ and $N'$ label the states of the heat bath and $Z$ is the partition function, $\mathcal{V} = \sum_i Q_i F_i$ couples system and bath, where $F_i$ and $Q_i$ operate on the system and bath, respectively. The reverse rate is $\gamma_{jk}= \gamma_{kj}e^{-\frac{E_j-E_k}{k_BT}}$. Although, the analytical expression is available, in practice, the rates are notoriously difficult to calculate as they require the knowledge of coupling term and the eigenspectrum  of the  bath. As a result the rates are often found in an empirical way, either using some judicious arguments \cite{q981-pd5j} or fitting to experimental data. Some examples include relaxation times for the Bloch equations \cite{BHATTACHARYYA202057}, relaxation parameter for stimulated emission and absorption \cite{Lin:10}, damping rates for the altermagnet-cavity system \cite{7bss-9yxb}.  In SM we derive the expressions for transition rates due to coupling between the polar phonon and acoustic phonons  in the framework of Debye model, which  under certain approximations,  predicts linear dependence of transition rates on the Bohr frequencies used here. However, since some of the parameters of the model are not readily available we follow the standard practice to fit rates to experimental data. In particular, the thermal average relaxation rate for our model is 
$\langle \gamma \rangle = 2 \sum_n p_n^{\mathrm{eq}} \gamma_n$, where $p_n^{\mathrm{eq}}$ is the Boltzmann probability of the eigenstate $n$. Then this value is fit to experimental $\gamma_{exp}$ for the corresponding mode through $k_{scale} = \gamma_{exp}/\langle \gamma \rangle$. The experimental values can be obtained from the literature on IR or Raman measurements, as they related to the width of the spectral lines. The values of $\gamma_{exp}$ for different materials used here are given in SM.

 {\it The role of intrinsic dynamics.} Including intrinsic dynamics increases the computational cost substantially and, in some cases, requires reducing the integration time step by a factor of 100. We attribute this to the presence of high Bohr frequencies, which reach up to 17 THz in \pzo. However, we expect the effects of intrinsic dynamics to average out on the timescale of relaxation. Indeed, $k_{scale}$ represents the ratio of the relaxation and intrinsic-dynamics timescales and, in our simulations, ranges from 4.8$\times$10$^{-3}$ to 6.6$\times$10$^{-3}$. To test this expectation, we repeated a hysteresis-loop simulation for \pzo\ at 300 K with intrinsic dynamics included and found no discernible differences in the resulting loop.

\begin{figure*}
\centering
\includegraphics[width=0.9\textwidth]{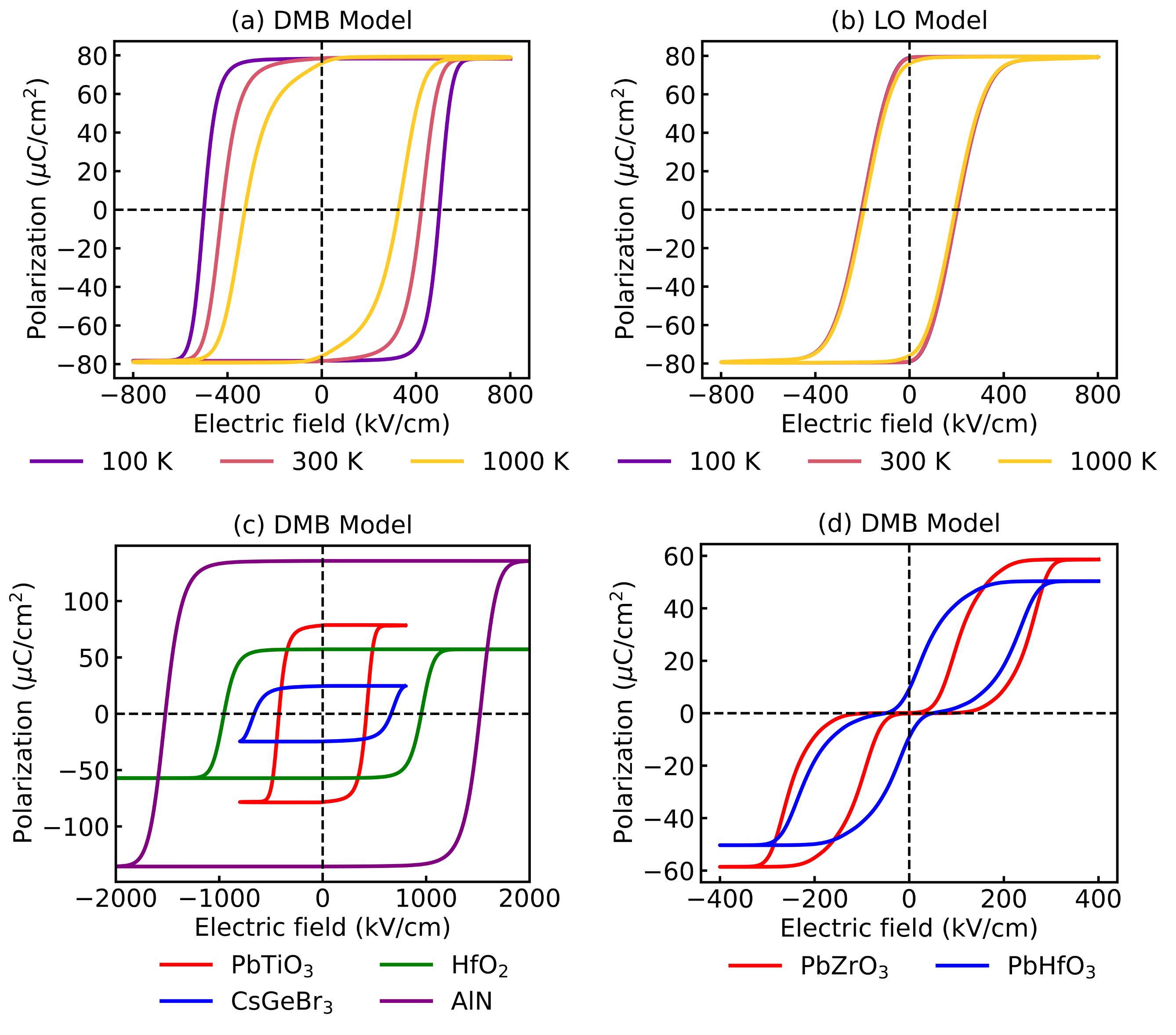}
\caption{(a)-(b) Polarization--electric field hysteresis loops of \pto\ computed at different temperatures using the models indicated in the panel titles. The simulations were performed using $\nu=1.7$ THz and an integration time step of $\Delta t=0.01$ fs. (c) Polarization--electric field hysteresis loops computed for different ferroelectric materials using the DMB model. The simulations were performed using $\nu=1.7$ THz and $\Delta t=0.01$ fs. (d) Polarization--electric field hysteresis loops computed for different antiferroelectric materials using the DMB model. The simulations were performed using $\nu=0.17$ THz and $\Delta t=0.1$ fs. All other simulation parameters are provided in the SM.}
\label{fig4}
\end{figure*}

{\it Application to ferroelectrics.} Figure~\ref{fig4}(a) and (b) shows hysteresis loops computed for ferroelectric \pto\ using the DFT-calculated $U(P)$ shown in Fig.~\ref{fig1}(a) and the two finite-temperature models developed in this work. The full set of results is provided in Fig.~S4. These results demonstrate that the developed approaches perform well for ferroelectrics, reproducing ferroelectric hysteresis loops.
We note that, for the LO model, the loops do not exhibit a significant temperature dependence at the chosen electric-field frequency. This may be due to the use of a temperature-independent relaxation rate.

To further assess the performance of the models, we apply them to additional ferroelectric and antiferroelectric materials, including nonperovskites. Figure~\ref{fig4}(c) shows hysteresis loops computed at 300 K for \pto, \cgb, \hfo, and \aln. The model reproduces the well-known trend that the coercive fields of \aln\ and \hfo\ are significantly larger than those of oxide ferroelectrics such as \pto. We also note that the coercive field predicted for \cgb\ is larger than that for \pto, which may be related to the use of a different exchange-correlation functional (see SM for details).

For \pho\ (Fig.~\ref{fig4}(d)) the model correctly predicts antiferroelectric loops. Experimental  fields for forward switching in \pho\ are in the range of 215--615 kV/cm and for backward switching they are in the range of 175--360 kV/cm~\cite{wei2019excellent,huang2021large,tsai2021antiferroelectric}. So our predictions are in in the range of experimental values. Note that we used the same AC-field frequency for all ferroelectric materials and the same AC-field frequency for all antiferroelectric materials. The frequencies differ between the two groups because DFT underestimates the coexistence field in \pzo\ and \pho. The coexistence field is defined as the electric field at which the polar and antipolar phases have the same energy and corresponds experimentally to the field at the center of the hysteresis loop. Within DFT, it is given by $E_{coex}=\Delta U/(VP)$, where $\Delta U$ is the zero-field energy difference between the polar and antipolar phases. We obtain coexistence fields of 181.2 and 124.9 kV/cm for \pzo\ and \pho, respectively, from DFT, compared to experimental values of 216.5 kV/cm~\cite{haddad2025chemical} and 190--470 kV/cm~\cite{wei2019excellent,huang2021large,tsai2021antiferroelectric}, respectively. This underestimation effectively contracts the hysteresis loops along the electric-field axis. To account for this effect, we use a lower AC-field frequency for the antiferroelectric materials.

Based on our results, we conclude that treating (anti)ferroelectric phase transitions as relaxational processes within a quantum mechanical framework offers an unconventional yet powerful approach to their modeling and understanding. This framework overcomes the limitations of the traditional description of such transitions as Arrhenius-type processes within a classical regime and enables efficient, first-principles–based simulations of phase transitions.
We anticipate that this approach will be applicable to a broad spectrum of other phase transitions -- including magnetic, ferroelastic, and multiferroic ones. Furthermore, the framework may provide a novel perspective for modeling chemical reaction rates, and applicable to modeling quantum tunneling phenomena, and electronic transitions.
Equally important, the model is directly applicable to phase transitions driven by electronic rather than ionic degrees of freedom  \cite{PhysRevLett.108.237601,doi:10.1021/acs.nanolett.4c00141,PhysRevB.104.195148,PhysRevResearch.3.L042028}, such as proton-ordering and hopping ferroelectrics \cite{doi:10.1021/jacs.0c02924}. Beyond these specific cases, the framework provides a direct means to probe quantum-mechanical features underlying a wide variety of phase transitions.

{\it Technical details of DFT simulations.}
All DFT calculations used to compute the energy profiles were performed using the Vienna Ab initio Simulation Package (VASP)~\cite{kresse1996efficient,kresse1996efficiency}. The projector augmented-wave (PAW) method~\cite{blochl1994projector,kresse1999ultrasoft} was employed to describe the ion--electron interaction. Exchange and correlation effects were treated within the local density approximation (LDA), except for \cgb, see SM for justification. A plane-wave kinetic-energy cutoff of 600 eV was used throughout.
Structural relaxations, where required, were performed using the conjugate-gradient algorithm with simultaneous relaxation of both the ionic coordinates and lattice parameters until the Hellmann--Feynman forces on all atoms were below 0.005 eV,\AA$^{-1}$. A $\Gamma$-centered $k$-point mesh corresponding to a reciprocal-space spacing of 0.22 \AA$^{-1}$ was used in all calculations.
Polarization was computed using the modern theory of polarization based on the Berry-phase formalism~\cite{king1993theory,vanderbilt1993electric,resta1994macroscopic}. Material-specific computational details are provided in the SM.

{\it Acknowledgments.}
N.M. and S.L. acknowledge financial support by the U.S. National Science Foundation under grant No. DMR-2219476.  I.P. acknowledges financial support by the U.S. Department of Energy, Office of Basic Energy Sciences, Division of Materials Sciences and Engineering under grant DE-SC0005245. Computational support was provided by the National Energy Research Scientific Computing Center (NERSC), a U.S. Department of Energy, Office of Science User Facility located at Lawrence Berkeley National Laboratory, operated under Contract No. DE-AC02-05CH11231 using NERSC award BES-ERCAP-0025236. L.J. acknowledges support from SFI grant SFI/21/US/3785. A.K. gratefully acknowledges support from Department of Education and Learning NI through grant USI-211. M.L. and N.B-G. gratefully acknowledge funding from the U.S. National Science Foundation grant DMR-2219476.  The authors thank Harrison Shirey for his help in testing codes. 
 
 {\it Data and codes availability.} All codes used to produce data are available in Ref.~\cite{ourgithub}. All data included in the paper and Supplementary Materials are available in Ref.~\cite{ourdata}.

\newpage
\onecolumngrid
\begin{center}
   \textbf{\Large Supplementary Material}
\end{center}

\renewcommand{\thefigure}{S\arabic{figure}}
\setcounter{figure}{0}

\section{Derivation of trace correcting term for Eq.(2)}
From Eq.(2) the change in the density matrix element due to relaxation is  
\[
\Delta \rho_{nm} \approx -\gamma_{nm} (\rho_{nm}-\delta_{nm}\rho_{nm}^{eq}) \Delta t
\]

\[
\rho_{nm} (t+\Delta t) = \rho'_{nm} = \rho_{nm} + \Delta \rho_{nm}
\]

After renormalization

\[
\rho''_{nm} = \frac{\rho'_{nm}}{\operatorname{Tr}\{\rho'\}}
\]

\[
\operatorname{Tr}\{\rho'\} = \operatorname{Tr}\{\rho + \Delta \rho \} = \operatorname{Tr}\{\rho\} + \operatorname{Tr}\{\Delta \rho\} = 1 + \operatorname{Tr}\{\Delta \rho\}
\]

\[
\rho''_{nm} = \frac{\rho'_{nm}}{1 + \operatorname{Tr}\{\Delta \rho\}} \approx \rho'_{nm} (1 - \operatorname{Tr}\{\Delta \rho\}) = (\rho_{nm} + \Delta \rho_{nm})(1 - \operatorname{Tr}\{\Delta \rho\})
\]

\[
= \rho_{nm} - \rho_{nm} \operatorname{Tr}\{\Delta \rho\} + \Delta \rho_{nm} - \Delta \rho_{nm} \operatorname{Tr}\{\Delta \rho\}
\]

We can neglect the last term in the above equation to get
\[
\rho''_{nm} - \rho_{nm} = \Delta \rho''_{nm} = \Delta \rho_{nm} - \rho_{nm} \operatorname{Tr}\{\Delta \rho\}
\]

\[
\Delta \rho''_{nm} = - \gamma_{nm} (\rho_{nm}-\delta_{nm}\rho_{nm}^{eq}) \Delta t - \rho_{nm} \sum_n\{- \gamma_{n0} (\rho_{nn}-\rho_{nn}^{eq}) \Delta t\}
\]
which can be reduced to the differential equation
\[
\frac{d \rho''_{nm}}{d t} = - \gamma_{nm} (\rho_{nm}-\delta_{nm}\rho_{nm}^{eq}) + \rho_{nm} \sum_j\gamma_{j0} (\rho_{jj}-\rho_{jj}^{eq})
\]

Let us now confirm that the equation preserves the trace of the density matrix.  First note, that $
\sum_n \gamma_{n0} (\rho_{nn}-\rho_{nn}^{eq}) $  is independent of subscripts $n$ and $m$ and therefore is the same for all density matrix elements. We call it const. 
Now lets compute the time derivative of the $\operatorname{Tr}\{\rho''\}$

\[
\frac{d}{d t}\sum_n \rho''_{nn} = \sum_n\bigl[- \gamma_{n0} (\rho_{nn}-\rho_{nn}^{eq})\bigr] + \text{const} \; \operatorname{Tr}\{\rho\}
\]

Since $\operatorname{Tr}\{\rho\} = 1$, the last term is just the previously defined const, leading to 

\[
\frac{d}{d t}\sum_n \rho''_{nn} = \sum_n\bigl[- \gamma_{n0} (\rho_{nn}-\rho_{nn}^{eq})\bigr] + \sum_n \gamma_{n0} (\rho_{nn}-\rho_{nn}^{eq}) = 0
\]

\section{Derivation of Relaxation rate}
Here we propose one possible route to compute relaxation rates following the approach of Ref~\cite{BlumKarl2011QToR}. 
\[
\gamma_{mn} = \frac{2\pi}{\hbar} \sum_{NN'} \left| \langle mN | \mathcal{V} | nN' \rangle \right|^2 
\, e^{-\beta E_{N'}} \delta(E_{N'} + E_N - \hbar\omega_{mn})
\]
Here $\mathcal{V} = \sum_i Q_i \, F_i $ is the term that couples the system $S$ and reservoir $R$, or heat bath, $N$ and $N'$ label the states of $R$, while $m$ and $n$ label the states of the $S$. In the coupling term $Q_i$ and $F_i$ are the operators acting on the $S$ and $R$, respectively.  We will use $\mathcal{V} = Q \, F $. $R$, for example, could be all other phonons in the $S$ and outside it with which it interacts.  Thus, 
\[
\langle m N| V | n N'\rangle= \langle m |  Q  | n \rangle \langle N |  F | N' \rangle\]
We now have
\[
\gamma_{mn} = \frac{2\pi}{\hbar Z}  \left| \langle m | Q | n \rangle \right|^2\sum_{N'} e^{-\beta E_{N'}} \sum_N \left| \langle N | F | N' \rangle \right|^2 
\,  \delta(E_{N'} - E_N - \hbar\omega_{mn})
\]
Next we denote $Q_{mn} = \langle m | Q | n \rangle$ and assume 
$\langle N | F | N' \rangle = \lambda =\text{constant} $, which results in 
 \[
\gamma_{mn} = |\lambda|^2 \frac{2\pi}{\hbar Z}  |Q_{mn}|^2 \sum_{N'} e^{-\beta E_N'}
\sum_N \delta(E_{N'} - E_N - \hbar\omega_{mn})
\]
The summation over $N$ is proportional to the number of states with energy $E_N=E_{N'}-\hbar \omega_{mn}$ which can be replaced with density of states for the heat bath $\rho(E_{N'}-\hbar\omega_{mn})$. We now have 

\[
\gamma_{mn} = \frac{|\lambda|^2 }{h} |Q_{mn}|^2
 \sum_{N'} \frac{e^{-\beta E_N}}{Z}
\rho(E_{N'} - \hbar \omega_{mn}) =  
 \frac{|\lambda|^2 }{h} |Q_{mn}|^2
\langle\rho(E_{N'} - \hbar \omega_{mn})\rangle 
\]
where brackets indicate the thermal average. Using again continuum energy states for $R$ we get 
\[
\langle\rho(E_{N} - \hbar \omega)\rangle = \frac{\int dE ~\rho(E) e^{-\beta E} \rho(E-\hbar \omega)}{\int dE ~\rho(E) e^{-\beta E} }
\]

Let us use Debye model for phonons so that the density of states is 
\[
\rho_D = \begin{cases} 
\frac{3V}{2\pi^2} \frac{\omega^2}{v_s^3} = \frac{3VE^2}{2\pi\hbar^3 v_s^3} = \mathcal{A} E^2 & \text{if } \omega < \omega_D = k_Dv_s ~or, E<\hbar \omega_D\\
0 & \text{otherwise} 
\end{cases}
\]
where $v_s$ and $V$ is the speed of sound and volume of the crystal, respectively. Since $R$ is much larger than $S$ we can use the approximation for $\hbar w_{mn}/E <<1$ to give
\[
\rho(E-\hbar \omega_{mn}) = \mathcal{A} (E-\hbar \omega_{mn})^2 \ \approx \mathcal{A}E^2 \left(1-\frac{2\hbar\omega_{mn}}{E}\right)
\]
Within this approximation we have 
\[
\langle\rho(E-\hbar \omega)\rangle = \mathcal{A} \left[\frac{\int dE ~E^4 e^{-\beta E}- 2\hbar \omega_{mn}\int dE ~E^3 e^{-\beta E}}{\int dE ~E^2 e^{-\beta E}} \right]
\]

Let us use substitution $\beta E = x$ to write the expression as follows

\[
\langle\rho(E-\hbar \omega)\rangle =  \frac{\mathcal{A}\left[ \frac{1}{\beta^5}\int_0^{x_D} dx ~x^4 e^{-x} -2\hbar \omega_{mn}\frac{1}{\beta^4}\int_0^{x_D} dx ~x^3 e^{-x}\right]}{\frac{1}{\beta^3}\int_0^{x_D} dx ~x^2 e^{-x}}
\]
Lets denote the definite integrals as
\[
\int_0^{x_D} dx ~x^4 e^{-x} = I_4,~~\int_0^{x_D} dx ~x^3 e^{-x} = I_3,~~\text{and} \int_0^{x_D} dx ~x^3 e^{-x} = I_2
\]

so that average density of states becomes

\[
\langle\rho(E-\hbar \omega)\rangle =  \mathcal{A} (k_BT)^2\frac{\left[ I_4 -\frac{2\hbar \omega_{mn}}{k_BT} I_3\right]}{I_2}
\]
Putting it all together produces

\[
\gamma_{mn} = \frac{|Q_{mn}|^2 |\lambda|^2 3V}{\hbar^4 v_s^3} (k_BT)^2 \left[ I_4' -\frac{2\hbar \omega_{mn}}{k_BT} I_3'\right]
\]

where $I_4'=\frac{I_4}{I_2}$ and $I_3'=\frac{I_3}{I_2}$.

\section{Technical details of calculations}



{\bf PbTiO$_3$:} The distortion path between the Pm$\bar{3}$m and P4mm phases was constructed using a single unit cell of PbTiO$_3$. For enenrgy landscape 64 of the u.c. were used to keep volume consistent for different materials. 

{\bf AlN:} To generate a structure with the opposite polarization direction, we applied a mirror-symmetry operation with a mirror plane perpendicular to the polar $c$ direction and passing through the Al site. The two structures with opposite polarization orientations were then connected using ISOTROPY to construct the distortion path, along which the energy and polarization were calculated.

We compute $Z^*=\sum_i Z_i^*\xi_i$, where $Z_i^*$ is the Born effective charge of ion $i$, computed from DFT, and $\xi_i$ are the components of the pseudoeigenvector. The pseudoeigenvector is defined as the normalized vector of ionic displacements connecting the nonpolar structure at the top of the energy barrier to the polar structure. In this case, $\xi_1(Al)=\xi_2(Al)=0$ and $\xi_1(N)=\xi_2(N)=1/\sqrt{2}$. The elemental Born effective charges are $Z^*_{Al}=2.67$ and $Z^*_N=-2.67$. This yields $Z^*=3.76e$.

{\bf HfO$_2$:} The distortion path was constructed by connecting the polar Pca2$_1$ structure to the cubic Fm$\bar{3}$m structure.

{\bf CsGeBr$_3$:} For DFT calculations, exchange and correlation effects were described using the r2SCAN functional~\cite{furness2020accurate}. LDA could not be used because it incorrectly predicts the cubic phase to be stable, in disagreement with experiment~\cite{furness2020accurate}. The distortion path was constructed by connecting the polar R3m structure to the cubic Pm$\bar{3}$m structure.

{\bf PbZrO$_3$:} The distortion path between the Pbam and R3c phases was constructed using a 64-unit-cell supercell of \pzo.

{\bf PbHfO$_3$:} To compute $Z^*$, we employed the same approach as for AlN, but constructed the pseudoeigenvector from the ionic displacements associated with the transition from the centrosymmetric Pm$\bar{3}$m phase to the polar R3c phase. Note that Pm$\bar{3}$m was used as the reference nonpolar phase instead of Pbam because the supercell required to connect Pbam and R3c is prohibitively large for Born effective charge calculations.

The pseudoeigenvector is $(0, 0.142, 0.316, 0.649, 0.649)$, where the components correspond to the displacements of Pb, Hf, and the three oxygen ions, respectively. The elemental Born effective charges are 3.94, 5.69, and $-3.21$ for Pb, Hf, and O, respectively. These values yield $Z^*=3.8e$ per f.u.

Note, that Born effective charge for a supercell is a product of the one for the polar unit cell and the number of such unit cells in the supercell. 
We could not find a suitable experimental reference for experimental relaxation rate of PbHfO$_3$ so we use the same as for \pzo.

\begin{table*}[!htbp]
\begin{center}
\caption{The model parameters (mass $M$, supecell volume $V$, polar mode frequency $\nu_{A_1}$, Number of formula units (No f.u.), experimental relaxation rate $\gamma_{exp}$ from the literature, scaling factor $k_{sacle}$ and Born effective charge $Z^*$ per polar unit cell)  for the materials considered here.}
\begin{tabular}{p{1.7cm} p{2cm}  p{2.2cm} p{1.7cm} p{1.9cm} p{1.5cm} p{2.5cm} p{1.5cm}} 
\hline 
Material & Mass (amu) & Volume (\AA$^3$) & $\nu_{A1}$ (THz) & No f.u. & $\gamma_{ex}$ (cm$^{-1}$) & k$_{scale}$ & Z$^*$ (e)\\
\hline 
\pzo\       & 17395  & 4407    & 3.14     & 64     & 43~\cite{ostapchuk2001polar}    & 4.556$\times$10$^{-2}$    & 6.32\cite{mani2015finite} \\
\pho\       & 13502  & 4313    & 2.61     & 64     & 43~\cite{ostapchuk2001polar}    & 4.506$\times$10$^{-2}$    & 3.80 \\
\pto\       & 5151   & 3760    & 4.18     & 64     & 32~\cite{sanjurjo1983pressure}    & 1.176$\times$10$^{-2}$    & 9.15\cite{mani2013atomistic} \\
\cgb\       & 1454   & 4762    & 6.0\cite{kashikar2026dft}      & 27     & 28~\cite{yan2025fully}    & 1.359$\times$10$^{-2}$    & 6.57\cite{kashikar2026dft} \\
\aln\       & 3727   & 4094   & 18.7     & 200     & 2.55~\cite{kazan2006temperature}  & 6.619$\times$10$^{-3}$   & 3.76 \\
\hfo\       & 117737 & 4446    & 3.81     & 144     & 7.2~\cite{li2009raman}   & 4.833$\times$10$^{-2}$    & 3.84\cite{kingsland2025first} \\
\hline
\end{tabular}
\label{tab:lattice-parameter}
\end{center}
\end{table*}

\begin{figure*}[!htbp]
\centering
\includegraphics[width=1.0\textwidth]{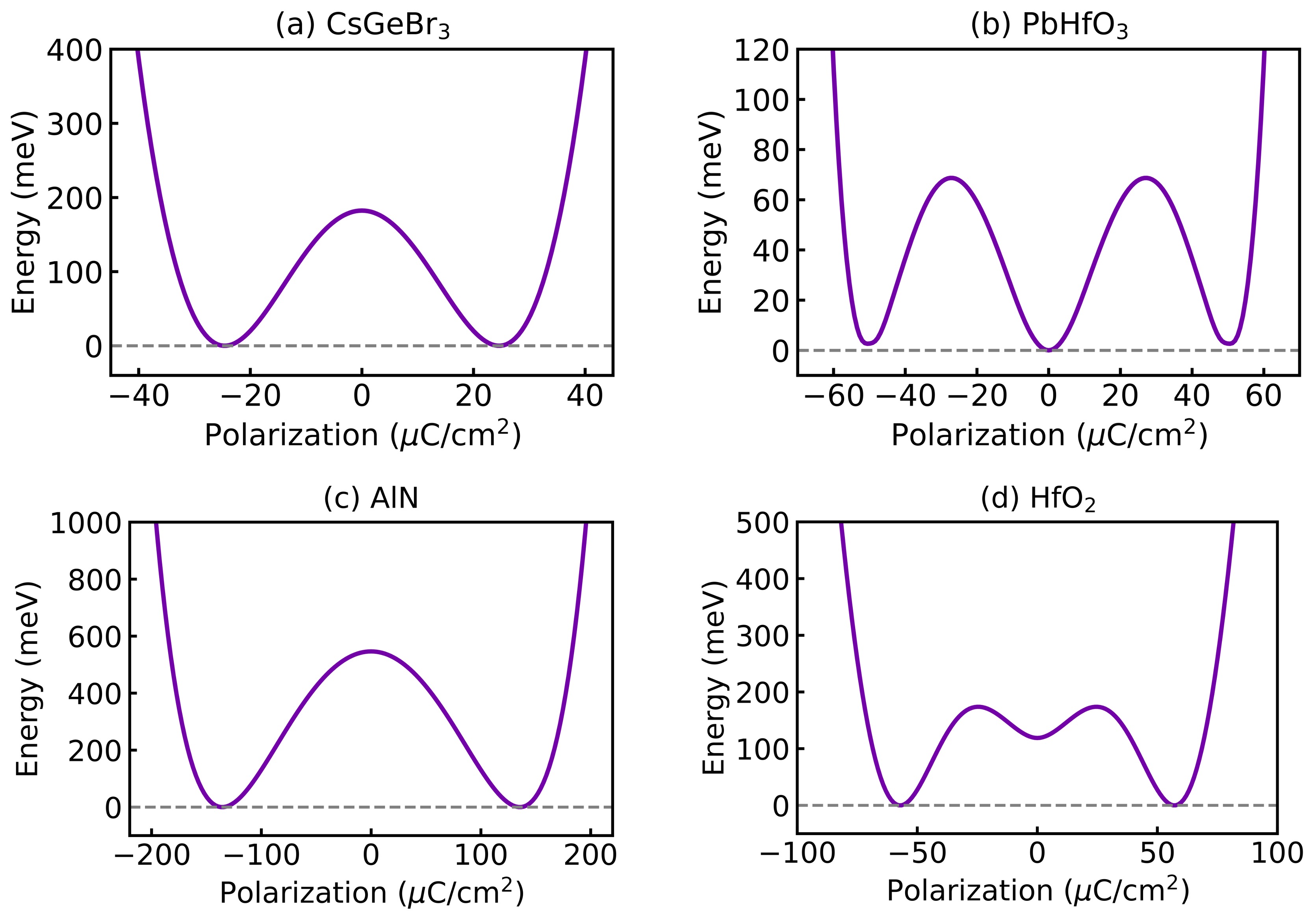}
\caption{Energy per formula unit as a function of polarization for ferroelectric PbTiO$_3$ (a), antiferroelectric PbHfO$_3$ (b),  ferroelectric AlN (d), and ferroelectric HfO$_2$ (d) as computed from DFT simulations.}
\label{Fig1}
\end{figure*}

\begin{figure*}[!htbp]
\centering
\includegraphics[width=1.0\textwidth]{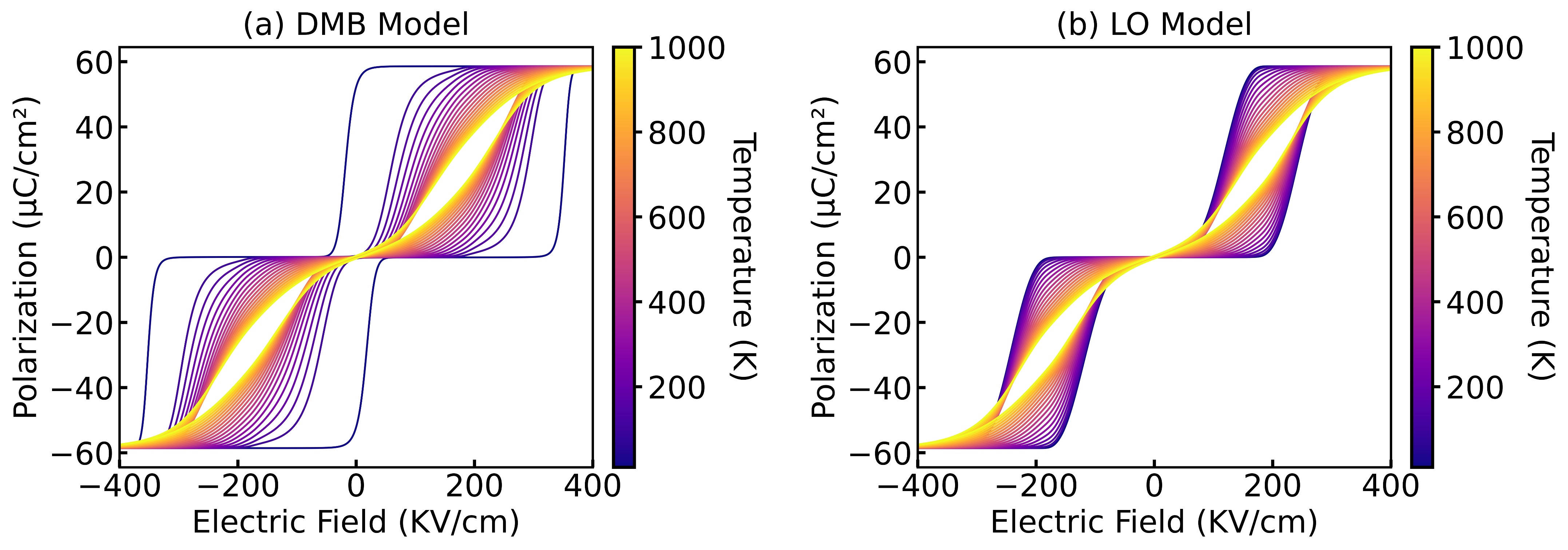}
\caption{(a)-(b) Polarization as a function of electric field in \pzo\ computed for different temperatures from 10 K to 1000 K using the models given in the titles.  The  following parameters were used: $M=$ 17395.53 amu, $\nu=$ 0.17 THz, $\Delta t=$~0.1~fs, $k_{scale}=$ 0.045.}
\label{Fig2}
\end{figure*}

\newpage

\begin{figure*}[!htbp]
\centering
\includegraphics[width=1.0\textwidth]{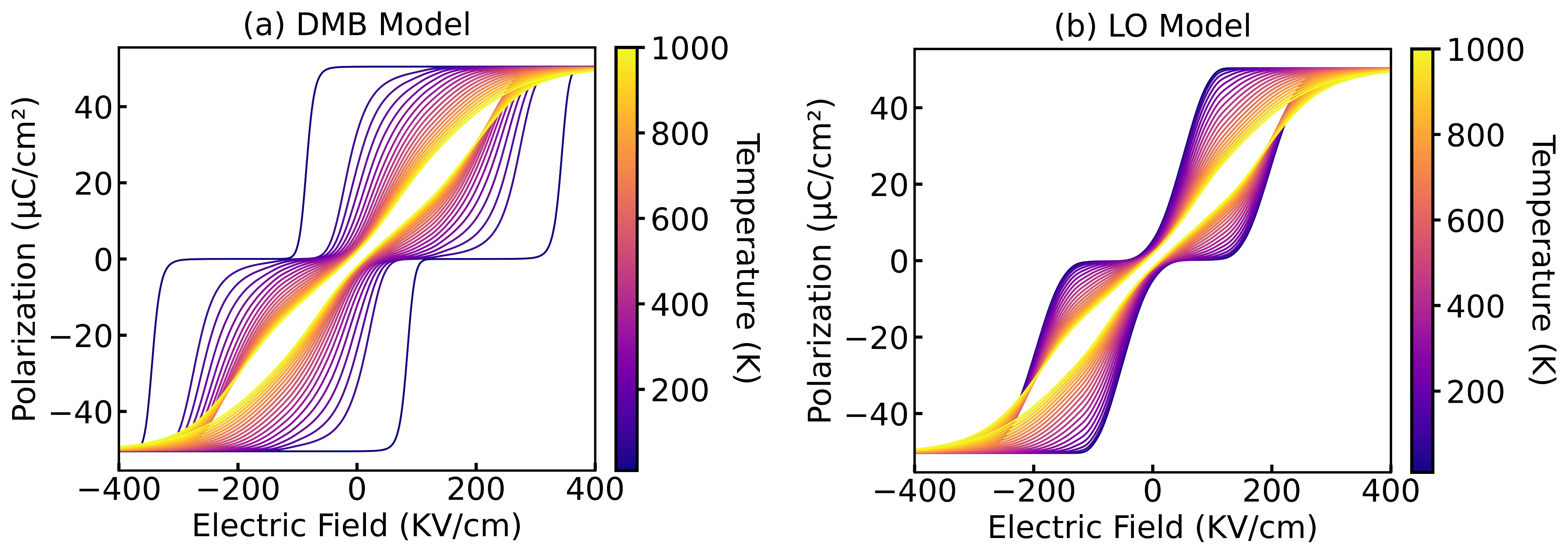}
\caption{(a)-(b) Polarization as a function of electric field in \pho\ computed for different temperatures from 10 K to 1000 K using the models given in the titles.  The  following parameters were used: $M=$ 13502.22 amu, $\nu=$ 0.17 THz, $\Delta t=$~0.1~fs, $k_{scale}=$ 0.045.}
\label{Fig3}
\end{figure*}

\begin{figure*}[!htbp]
\centering
\includegraphics[width=1.0\textwidth]{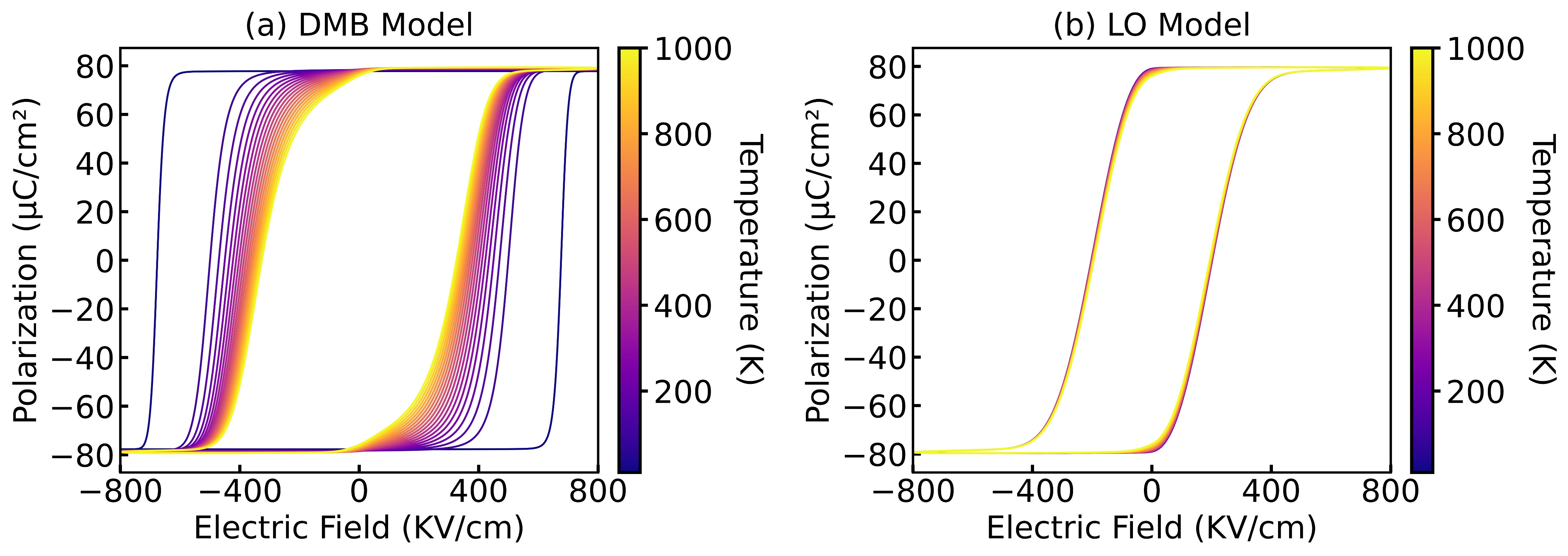}
\caption{(a)-(b) Polarization as a function of electric field in \pto\ computed for different temperatures from 10 K to 1000 K using the models given in the titles.  The  following parameters were used: $M=$ 5151.09 amu, $\nu=$ 1.7 THz, $\Delta t=$~0.01~fs, $k_{scale}=$ 0.012.}
\label{Fig4}
\end{figure*}

\begin{figure*}[!htbp]
\centering
\includegraphics[width=1.0\textwidth]{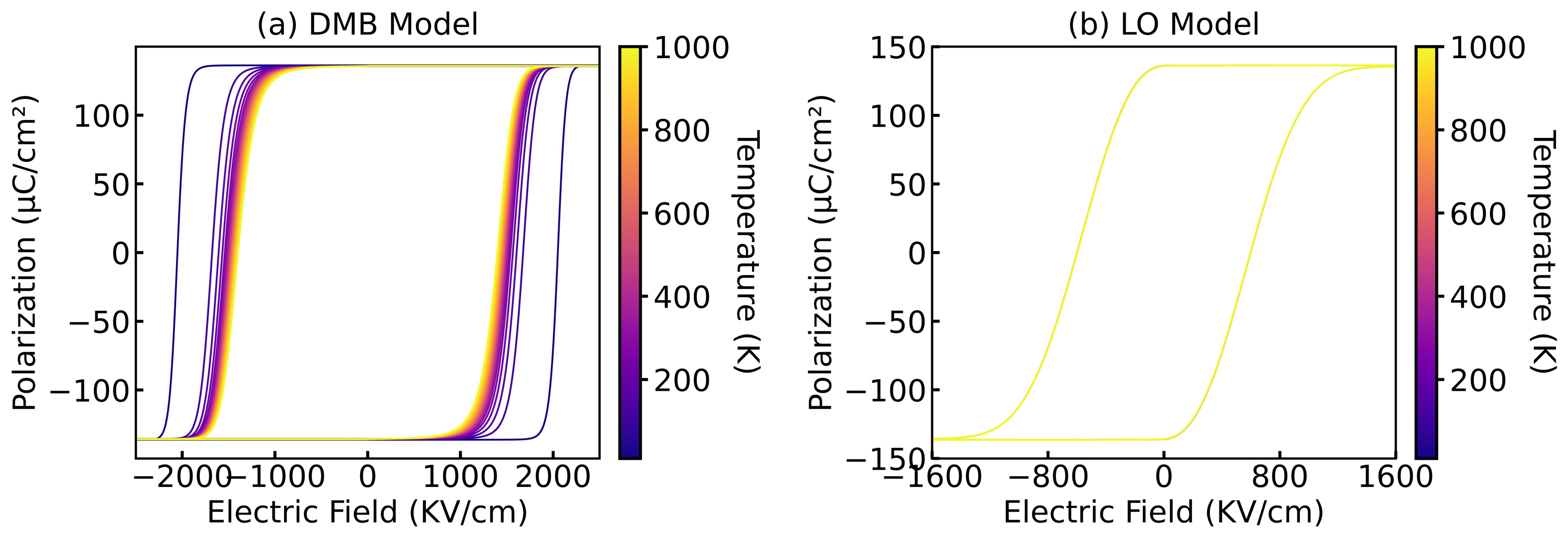}
\caption{(a)-(b) Polarization as a function of electric field in \cgb\ computed for different temperatures from 10 K to 1000 K using the models given in the titles.  The  following parameters were used: $M=$ 1454.27 amu, $\nu=$ 1.7 THz, $\Delta t=$~0.01~fs, $k_{scale}=$ 0.013.}
\label{Fig5}
\end{figure*}

\begin{figure*}[!htbp]
\centering
\includegraphics[width=1.0\textwidth]{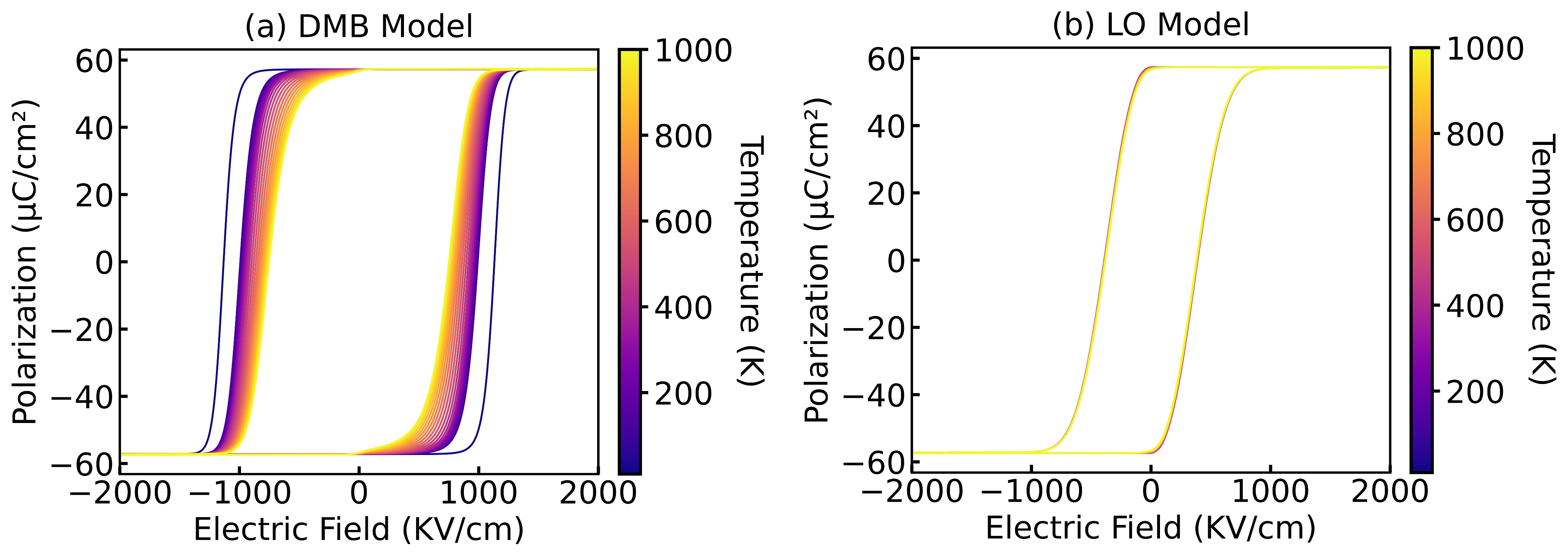}
\caption{(a)-(b) Polarization as a function of electric field in \aln\ computed for different temperatures from 10 K to 1000 K using the models given in the titles.  The  following parameters were used: $M=$ 3727.38 amu, $\nu=$ 1.7 THz, $\Delta t=$~0.01~fs, $k_{scale}=$ 0.007.}
\label{Fig6}
\end{figure*}

\begin{figure*}[!htbp]
\centering
\includegraphics[width=1.0\textwidth]{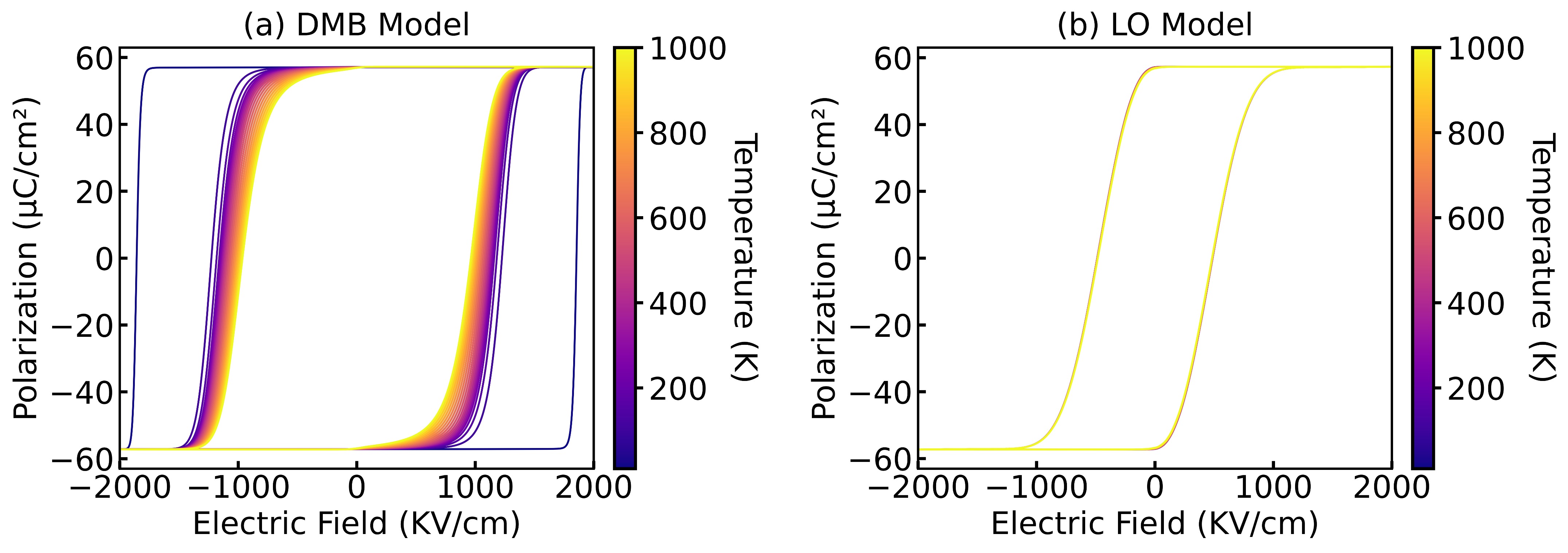}
\caption{((a)-(b) Polarization as a function of electric field in \hfo\ computed for different temperatures from 10 K to 1000 K using the models given in the titles.  The  following parameters were used: $M=$ 117737.86 amu, $\nu=$ 1.7 THz, $\Delta t=$~0.01~fs, $k_{scale}=$ 0.018.}
\label{Fig7}
\end{figure*}

\clearpage
\phantomsection
\addcontentsline{toc}{section}{References} 


\end{document}